\newcommand{\tit}[1]{``#1,''}
\newcommand{\C}{{\mathbb C}}
\newcommand{\Hil}{{\mathcal H}}
\newcommand{\R}{{\mathbb R}}
\newcommand{\Cl}{{\mathfrak{Cl}}}
\newcommand{\Id}{{\mathbb I}}
\newcommand{\G}{{\mathbf\Gamma}}
\newcommand{\un}{{\mathfrak u}}
\newcommand{\gl}{{\mathfrak{gl}}}
\newcommand{\so}{{\mathfrak{so}}}
\newcommand{\mbf}[1]{\mbox{\boldmath$#1$}}
\newcommand{\bsgm}[1]{\mbf{\sigma}_\mathsf{#1}}
\newcommand{\btau}[1]{\mbf{\tau}_\mathsf{#1}}
\newcommand{\eq}[1]{{\rm Eq.~(\ref{#1})}}

\documentclass{article}
\usepackage{amsfonts}
\begin{document}
\title{Clifford algebras, noncommutative tori and universal quantum computers}
\author{Alexander Yu. Vlasov\thanks{
 E-mail: {\tt qubeat@mail.ru} or {\tt alex@protection.spb.su}}\\
\small Federal Radiological Center (IRH),
197101, Mira Street 8, St.-Petersburg, Russia}
\date{14.06.2001 (arXiv: quant-ph/0109010 03.09.2001)}
\maketitle
\begin{abstract}
Recently author suggested \cite{VlaUCl} an application of
Clifford algebras for construction of a ``compiler'' for universal binary
quantum computer together with later development \cite{VlaUNt}
of the similar idea for a non-binary base. The non-binary case is related
with application of some extension of idea of Clifford algebras. It is
noncommutative torus defined by polynomial algebraic relations of order $l$.
For $l=2$ it coincides with definition of Clifford algebra. Here is presented
the joint consideration and comparison of both cases together with some
discussion on possible physical consequences.
\end{abstract}
\sloppy

\section{Introduction}

Application of Clifford algebras \cite{ClDir,Post} in theory of quantum
computers \cite{PB82,FeySim} already was discussed in relation with product
operator formalism \cite{BR94,SCH98} used for description of NMR based
quantum computers \cite{SCH98,HD0}. In the present paper is discussed
application of Clifford algebras in more abstract models of quantum
computations for constructions of universal sets of quantum gates.
It is related already with early works about quantum computers, as
``universal simulator'' of behavior of quantum systems
\cite{FeySim,FeyComp,DeuTur,DeuGate}.

Due to such approach it is reasonable to consider some sets of elements
that may be used for construction or approximation of {\em any} unitary
transformation, like {\sf AND}, {\sf NOT} gates for usual Boolean
function used in classical computations.

In most common model of quantum computation each elementary system is
described by finite-dimensional Hilbert space and quantum gates act on some
composite system with Hilbert space described as tensor product.

In simplest case the elementary system is described by two-dimensional
Hilbert space and called {\em qubit}. More general case with
higher-dimensional systems \cite{Gott98} is used less often, but also is
considered in the paper. For quantum computation with $n$ qubits total
Hilbert space has dimension $2^n$ and most general quantum evolution
may be described by $2^n \times 2^n$ unitary matrix.

In such a case the problem of universality is existence of some set
of unitary matrices $U_\kappa$ with possibility of presentation of any
unitary matrix $U \in {\rm U}(2^n)$ as product
\begin{equation}
U = U_{\kappa_1} U_{\kappa_2} \cdots U_{\kappa_p}
\label{UnG}
\end{equation}
or approximation $U \approx U_{k_1} U_{k_2} \cdots U_{k_p}$
with necessary or arbitrary precision, if the set is finite
\mbox{$k = 1,\ldots,K$}.

It should be mentioned also that the Hilbert space $\Hil$ has special
structure due to representation as tensor power of $n$ two-dimensional
spaces $\Hil_2$
\begin{equation}
 \Hil = \underbrace{\Hil_2\otimes\cdots\otimes\Hil_2}_n .
\label{Hil2prod}
\end{equation}
 Due to such structure any unitary transformation of
some subsystem with $m < n$ qubits, $U_S \in {\rm U}(2^m)$ can be
enclosed in the ${\rm U}(2^n)$ as transformation that acts only on $m$
terms in the Hilbert space decomposition as tensor product with $n$ terms
\eq{Hil2prod}. Such transformations are called $m$-qubit gates.

The consideration has straightforward generalization on systems with dimension
of Hilbert space $\Hil_l$, $l>2$, with $l^n$-dimensional space of composition
\begin{equation}
 \Hil = \underbrace{\Hil_l\otimes\cdots\otimes\Hil_l}_n ,
\label{Hilprod}
\end{equation}
and unitary group U$(l^n)$. Non-binary quantum $m$-gates correspond to
elements of U$(l^m)$, $m \leq n$.

An important result of quantum theory of computation is existence
of universal set with two-qubit quantum gates \cite{DV95,Gates95,DeuUn}.

An application of Clifford algebras to quantum circuits with qubit is
justified even from formal point of view, for example universal complex
Clifford algebra $\Cl(2n,\C)$ is isomorphic with $\C(2^n\times 2^n)$, algebra
of all $2^n \times 2^n$ complex matrices \cite{ClDir}, but any quantum
gate may be represented as unitary $2^n \times 2^n$ matrix and considered as
element of the Clifford algebra $\Cl(2n,\C)$.

Less formal relation of Clifford algebras with quantum computation
may be also due to physical properties of system used for implementation
of quantum computer. For example property of {\em spin group} is important
in realization of quantum computers on NMR \cite{SCH98,HD0} systems.
Secondary quantization of fermions using Clifford systems \cite{WeylGQM,AQFT}
is also may be applied to quantum computation \cite{VlaTmr,VlaBak}.

In many papers about quantum computation is used presentation of quantum
gates in product operator formalism with Pauli matrices similar with one of
constructions of Clifford algebras \cite{ClDir}. The relation is directly
mentioned and used by some authors \cite{SCH98,HD0,VlaTmr}.
It should be mentioned that in \cite{VlaTmr,VlaUCl} and in present paper
are used complex Clifford algebras instead of real \cite{SCH98,HD0}.

\section{Universal sets of quantum gates}

In present paper is discussed application of Clifford algebras to
construction of universal set of quantum gates. It has some difference with
example discussed above, because instead of representation of unitary group
of quantum gates here is used representation of Lie algebra of this group.
It was suggested in \cite{VlaUCl} for quantum gates with qubits, and
later was found that for non-binary case very similar construction
can be used with some analogue of Clifford algebra, noncommutative
torus \cite{VlaUNt}. Here both cases are considered jointly, to
show similarity of constructions with noncommutative tori and Clifford
algebras together with some specific differences.

\medskip

Algebraic tools discussed here produce directs algorithms for decomposition
of some unitary matrix on set of quantum gates \cite{VlaUCl,VlaUNt} (sometime
such algorithms are called ``quantum compilers'' \cite{arudi}). It also can
be convenient as a background for design of quantum processors
\cite{VlaCla,VlaUQP}.

\subsection{Lie algebras}

Here is discussed infinitesimal approach to construction of universal
set of quantum gates using Lie algebra $\un(N)$. Here $N=2^n$ for
quantum computation with $n$ qubits. It was already discussed
in many works \cite{DV95,DeuUn,UnSim} and described here only briefly.
In the approach is considered Lie algebra $\un(N)$ of Lie group of unitary
matrices and finite set of elements $A_k$ of this algebra
represented as some anti-Hermitian matrices $A_k = -A_k^\dag$.
Quantum gates are generated as
\begin{equation}
U_k^\tau = \exp(A_k \tau)
\label{UnA}
\end{equation}
with some real parameter $\tau$. Now instead of universal set of elements
of Lie group U$(N)$ used in \eq{UnG} it is possible to work with elements
$A_k$ of Lie algebra $\un(N)$.

{\em If the elements $A_k$ generate full Lie algebra $\un(N)$ by
commutators, then it is possible to use set of gates described by \eq{UnA}
as universal set of gates} \cite{DV95,DeuUn,UnSim}.

The passage from Lie algebra to Lie group of gates due to \eq{UnA} is not
discussed with more detail, see \cite{DeuGate,DV95,DeuUn,UnSim,VlaUCl,VlaUNt}.
In physical applications $A_k = i H_k$ correspond to Hamiltonians and real
parameter $\tau$, may be infinitesimal \cite{DV95}, irrational \cite{DeuUn}
or it is possible to consider \eq{UnA} as one-parametric family, where $\tau$
is time \cite{UnSim}, but the main theme of this work is construction
of the set of elements $A_k$ itself using Clifford algebras and noncommutative
tori.

\subsection{Clifford algebras}

Any associative algebra is Lie algebra in respect to commutator
$[a,b]=ab-ba$. It was already mentioned that universal Clifford
algebra $\Cl(2n,\C)$ is isomorphic with associative matrix algebra
$\C(2^n \times 2^n)$. Lie algebra $\un(2^n)$ can be represented
as algebra of anti-Hermitian $2^n \times 2^n$ matrices in respect to
commutator and so can be considered as subalgebra of $\Cl(2n,\C)$.

Let us denote Pauli matrices
$\bsgm x, \bsgm y, \bsgm z \in \C(2\times 2) \cong \Cl(2,\C)$.
Then anti-Hermitian matrices
\begin{eqnarray}
 \G_{2k} & = &
 i \underbrace{\Id\otimes\cdots\otimes\Id}_{n-k-1}\otimes
 \bsgm x\otimes\underbrace{\bsgm z\otimes\cdots\otimes\bsgm z}_k \, ,
 \nonumber\\
 \G_{2k+1} & = &
 i \underbrace{\Id\otimes\cdots\otimes\Id}_{n-k-1}\otimes
 \bsgm y\otimes\underbrace{\bsgm z\otimes\cdots\otimes\bsgm z}_k \, ,
 \label{defG}
\end{eqnarray}
where $k = 0,\ldots,n-1$
are generators of $\Cl(2n,\C)$ \cite{ClDir,WeylGQM,AQFT,VlaUCl}.
The generators meet to usual rule:
\begin{equation}
 \{\G_k,\G_j\} \equiv \G_k \G_j + \G_j \G_k = -2\delta_{kj}
\label{acomG}
\end{equation}
where unit of algebra as multiplier is omitted for simplicity in right
part of this equation.

Sums of all possible {\em products} of the elements \eq{defG} with
{\em complex} coefficients generate full algebra $\Cl(2n,\C)$, but for
applications with Lie algebras and quantum gates we may use only
{\em commutators} with {\em real} coefficients. It can be simply
shown \cite{VlaUCl} that elements \eq{defG} generate only
$(2n^2+n)$-dimensional subalgebra in $4^n$-dimensional Lie algebra $\un(2^n)$
and the algebra is isomorphic with Lie algebra $\so(2n+1)$ and the
subgroup of U$(2^n)$ generated by the elements \eq{defG} is somorphic with
$(2n^2+n)$-dimensional group Spin$(2n+1)$. So the gates are not universal.

But simple commutation laws of Clifford algebra make possible to prove
that it is enough to join only one element $\G_u$ to $2n$ elements
\eq{defG} and the new set with $2n+1$ elements may generate full
$4^n$-dimensional Lie algebra $\un(2^n)$ \cite{VlaUCl}. As the extra element
may be used product of any three or four matrices $\G_k$, for example
\begin{equation}
\G_u = i\,\G_0\G_1\G_2
\label{defGu}
\end{equation}
or
\begin{equation}
\G_{u'} = i\,\G_0\G_1\G_2\G_3 .
\end{equation}
Here $\G_u$ and $\G_{u'}$ are also anti-Hermitian and so commutators and sums
with real coefficients may generate only anti-Hermitian matrices.

The set with $2n+1$ elements \eq{defG} and \eq{defGu} is universal, but
elements $\G_{2k}$ and $\G_{2k+1}$ correspond to $(k{+}1)$-qubit gates. It
is not very difficult to found new set there $2n+1$ elements are only one- and
two-qubit gates \cite{VlaUCl}. It is enough instead of \eq{defG} to consider
$\G_0$ together with $2n-1$ elements $\G_k\G_{k+1}$, $k=0,\ldots,2n-2$.
All elements of the new set generate only one- or two-qubit gates
because such products of two elements \eq{defG} have no more than two
non-unit terms in tensor decomposition (other terms are
$\bsgm z \bsgm z = \Id$),
but it is possible to produce all initial matrices $\G_k$ using the new set
\cite{VlaUCl}.

On the other hand, the initial set with elements \eq{defG} may be more
convenient for representation with fermionic annihilation and creation
operators \cite{ClDir,WeylGQM,AQFT} used in applications for quantum
circuits \cite{VlaTmr,VlaBak,VlaUCl}.

\subsection{Noncommutative tori}

The construction with Clifford algebra described above may not be
applied directly for non-binary quantum circuits, when each system
described by $l$-dimensional Hilbert spaces, $l>2$, and operators
from U$(l^n)$ are represented in tensor product of $n$ complex matrices
$\C(l \times l)$. But even in such a case the general principles are
very close \cite{VlaUNt}.

First, let us consider instead of Pauli matrices Cayley-Weyl-Connes
pair \cite{WeylGQM,ConnesNG}:
\begin{equation}
 U V = \exp(2 \pi i / l) V U,\quad V V^\dag = U U^\dag = \Id,
\label{T2Def}
\end{equation}
where $l \times l$ complex matrices $U$ and $V$ may be represented as
\begin{equation}
 U_{kj} = \delta_{k+1 ({\rm mod}~l),j}\, , \quad
 V_{kj} = \exp(2 \pi i k/l) \delta_{kj}.
\label{defUV}
\end{equation}
Lately it is called sometime ``general Pauli group'' \cite{Gott98}.

Let us denote $\zeta = \exp(2 \pi i / l)$. It is possible to consider
three elements
\begin{equation}
\btau x = U, \quad
\btau y = \zeta^{(l-1)/2}U V, \quad
\btau z = V,
\end{equation}
with properties \cite{VlaUNt}
\begin{equation}
 \btau x \btau y = \zeta \btau y \btau x,\quad
 \btau y \btau z = \zeta \btau z \btau y,\quad
 \btau x \btau z = \zeta \btau z \btau x,\quad
 \btau\mu^l = \Id.
\label{taucom}
\end{equation}
It is clear, that for $l=2$ \eq{taucom} coincide with anti-commutation
relation for Pauli matrices, but for $l>2$ here is some asymmetry
or {\em order}, because $\zeta^{-1} \ne \zeta$,
$\btau z \btau x = \zeta^{-1} \btau x \btau z \ne \zeta \btau x \btau z$
and it is not possible to treat all three elements $\btau\mu$ in equal way.
On the other hand, the order makes possible to introduce all necessary
construction with noncommutative torus quite simply \cite{VlaUNt}.

Let us introduce set of $\C(l^n \times l^n)$ matrices similar with
construction \eq{defG} above:
\begin{eqnarray}
 {\bf T}_{2k} & = &
  \underbrace{\Id\otimes\cdots\otimes\Id}_{n-k-1}\otimes
 \btau x\otimes\underbrace{\btau z\otimes\cdots\otimes\btau z}_k \, ,
 \nonumber\\
 {\bf T}_{2k+1} & = &
  \underbrace{\Id\otimes\cdots\otimes\Id}_{n-k-1}\otimes
 \btau y\otimes\underbrace{\btau z\otimes\cdots\otimes\btau z}_k \, ,
 \label{defT}
\end{eqnarray}
where $k = 0,\ldots,n-1$.
Then (see Ref.~\cite{VlaUNt}) $l^{2n}$ different products of ${\bf T}_k$
$k = 0,\dots,2n-1$ generate algebra of complex matrices $\C(l^n \times l^n)$
and the $2n$ elements ${\bf T}_k$ itself meet to special rule:
\begin{equation}
{\bf T}_j {\bf T}_k = \zeta {\bf T}_k {\bf T}_j ~~ (j < k),
\quad ({\bf T}_k)^l = \Id.
\label{TorDef}
\end{equation}
The \eq{TorDef} coincides with definition of Clifford algebra
for $l=2$, and for more general case $l > 2$ it is called here
{\em noncommutative torus}.\footnote{The term is used for example in
\cite{ConnesNG} for $n=2$ and irrational noncommutative torus, $l \in \R$,
a more general case with $n \ge 2$ is also widely known on some other areas
of research, but here is important to emphasize for comparison that using
same number $\zeta$ in \eq{TorDef} instead of matrix $\zeta_{kj}$ is
important for the application in present work}

Really matrices \eq{defT} are neither anti-Hermitian, nor Hermitian,
and for $l>2$ they generate complex Lie algebra $\gl(l^n,\C)$ \cite{VlaUNt}
without necessity of some extra elements like \eq{defGu} in case of Clifford
algebra. To produce generators of Lie algebra $\un(l^n)$ it is necessary
to use instead of each matrix ${\bf T}_k$ \eq{defT} two anti-Hermitian
matrices
\begin{equation}
 {\bf T}^+_k = i ({\bf T}_k + {\bf T}^\dag_k), \quad
{\bf T}^-_k = ({\bf T}_k - {\bf T}^\dag_k).
\label{Hpm}
\end{equation}

Here is also possible to introduce set of two-gates. It is again an analogy of
case with qubits and Clifford algebras discussed above \cite{VlaUCl,VlaUNt},
and instead of elements \eq{defT} it is possible to use ${\bf T}_0$ together
with $2n-1$ elements ${\bf T}^\dag_k{\bf T}_{k+1}$, $k=0,\ldots,2n-2$.
Such products of two elements \eq{defT} have no more than two non-unit terms
in decomposition (because $\btau z^\dag \btau z = \Id$), but the new set may
generate all matrices ${\bf T}_k$ and so equal with initial set. It is
possible to produce two anti-Hermitian matrices for each elements of the new
set using two sums like in \eq{Hpm}, the matrices correspond to one- or
two-gates \cite{VlaUNt}.

\section{Comparison and discussion}

It is clear from consideration above that using Clifford algebras not only
produced some alternative method for construction of universal set of
quantum gates for binary quantum circuits, but also gave some guidelines
for non-binary case. For example, expression for construction of
noncommutative tori as tensor product of Weyl pairs \eq{defT} corresponds
to \eq{defG} with generators of Clifford algebra and Pauli matrices.

The presentation of two-gates as product of two generators discussed
above is also similar for both cases.
It should be mentioned also that the construction has some interesting
physical consequence even outside of area of quantum computation,
because it is a demonstration, that evolution of set of few quantum
systems\footnote{described by equal finite-dimensional Hilbert spaces,
but without taking (anti)symmetrization into account} may be represented
using only pairwise interactions \cite{Grib0}.

On the other hand there are some differences between binary case $l=2$ with
qubits and Clifford algebras and non-binary case with noncommutative tori for
$l>2$. For example, generators of noncommutative tori are unitary like
Pauli matrices, but for $l>2$ they are not Hermitian and for generation of
Lie algebra $\un$ it is necessary to use two sets of generators \eq{Hpm}
instead of simple multiplication on imaginary unit in \eq{defG}. But for case
with Clifford algebra the \eq{defG} was also not enough, and it was necessary
to use extra one element like \eq{defGu}.

The extra gate with product of three or four generators may be quite essential,
as it already was emphasized in \cite{VlaUCl}. If there is some algebra with
an associative product, it is simple to introduce a structure of Lie algebra
using commutator, but opposite task is not straightforward. If there is some
Lie algebra with a ``bracket operation $[\cdot,\cdot]$'' satisfying necessary
axioms, then ``restoring'' of the product is related with construction of
universal enveloping algebra \cite{Post}.

For example under consideration the initial Lie algebra generated by
$2n$ elements \eq{defG} was simply Lie algebra of rotation group of
$(2n+1)$-dimensional space and expression like product with three generators
\eq{defGu} corresponds to consideration of complex enveloping algebra
isomorphic to $\C(2^n \times 2^n)$ with exponentially larger dimension.
If these constructions are not only abstract ones, but also correspond to
some structures and symmetries of a particular physical model used for
implementation of quantum computer, the extra one element like $\G_u$ may
have special physical entity, but it should be discussed elsewhere.


\begin{thebibliography}{99}
 \bibitem{ClDir} J. E. Gilbert and M. A. M. Murray, {\em Clifford algebras and
  Dirac operators in harmonic analysis}, (Cambridge University Press,
  Cambridge 1991).
 \bibitem{Post} M. M. Postnikov, {\em Lie groups and Lie algebras},
  (Nauka, Moscow 1982).
 \bibitem{WeylGQM} H. Weyl, {\em The Theory of Groups and Quantum
  Mechanics}, (Dover Publications, New York 1931).
 \bibitem{AQFT} J. C. Baez, I. E. Segal, Z. Zhou, {\em Introduction to
  Algebraic and Constructive Quantum Field Theory}, (Princeton University
  Press, Princeton 1992).
 \bibitem{ConnesNG} A. Connes, {\em Noncommutative Geometry},
  (Academic Press, San Diego 1994).
 \bibitem{BR94} B. Boulat and M. Rance, \tit{Algebraic formulation of the
  product operator formalism in the numerical simulation of the dynamic
  behaviour of multispin systems} {\em Mol. Phys. \bf 83} 1021 (1994). 
 \bibitem{SCH98} S. S. Somaroo, D. G. Cory, T. F. Havel, \tit{Expressing the
  operations of quantum computing in multiparticle geometric algebra}
  {\em Phys. Lett. \bf A 240} 1 (1998). 
 \bibitem{HD0} T. F. Havel and C. J. L. Doran, \tit{Geometric algebra
  in quantum information processing}  {\em Preprint} {\tt arXiv:quant-ph/0004031} (2000).
 \bibitem{PB82} P. Benioff, \tit{Quantum Mechanical Hamiltonian
  Models of Discrete Processes That Erase Their Own Histories: Application
  to Turing Machines} {\em Int. J. Theor. Phys. \bf 21}, 177 (1982). 
 \bibitem{FeySim} R. Feynman, \tit{Simulating Physics with
  Computers} {\em Int. J. Theor. Phys. \bf 21} 467 (1982).
 \bibitem{FeyComp} R. Feynman, \tit{Quantum-Mechanical
  Computers} {\em Found. Phys. \bf 16} 507 (1986).
 \bibitem{DeuTur} D. Deutsch, \tit{Quantum theory, the Church-Turing
  principle and the universal quantum computer}
  {\em Proc. R. Soc. London Ser. A \bf 400}, 97 (1985).
 \bibitem{DeuGate} D. Deutsch, \tit{Quantum computational networks}
  {\em Proc. R. Soc. London Ser. A \bf 425}, 73 (1989).
 \bibitem{DV95} D. P. DiVincenzo,
 \tit{Two-bit gates are universal for quantum computation}
  {\em Phys. Rev. \bf A 51}, 1015 (1995).
 \bibitem{Gates95} A. Barenco, C. H. Bennett, R. Cleve, D. P. DiVincenzo,
  N. Margolus, P. W. Shor, T. Sleator, J. A. Smolin, and H. Weinfurter,
 \tit{Elementary gates for quantum computation}
 {\em Phys. Rev. \bf A  52}, 3457 (1995). 
 \bibitem{DeuUn} D. Deutsch, A. Barenco, and A. Ekert,
 \tit{Universality in quantum computation}
  {\em Proc. R. Soc. London Ser. A \bf 449}, 669 (1995).
 \bibitem{UnSim} S. Lloyd, \tit{Universal quantum simulators}
  {\em Science \bf 273}, 1073 (1996).
 \bibitem{Gott98} D. Gottesman, \tit{Fault-tolerant quantum computation
  with higher-dimensional systems} {\em Lect. Not. Comp. Sci. \bf 1509},
  302 (1999); {\em Preprint} {\tt arXiv:quant-ph/9802007}.
 \bibitem{arudi} R. R. Tucci, \tit{A rudimentary quantum compiler}
  {\em Preprint} {\tt arXiv:quant-ph/9805015}, (1998).
 \bibitem{Grib0} A. A. Grib,~ personal communications.
 \bibitem{VlaTmr} A. Yu. Vlasov, \tit{Quantum gates and Clifford algebras}
  {\em Preprint} {\tt arXiv:quant-ph/9907079}, (1999).
 \bibitem{VlaBak} A. Yu. Vlasov, \tit{Algebra, logic and qubits: quantum
  abacus} {\em Preprint} {\tt arXiv:quant-ph/0001100}, (2000).
 \bibitem{VlaUCl} A. Yu. Vlasov,  \tit{Clifford algebras and universal
  sets of quantum gates}
  {\em Preprint} {\tt arXiv:quant-ph/0010071} (2000);
  {\em Phys. Rev. \bf A} 63, 054302 (2001).
 \bibitem{VlaUNt} A. Yu. Vlasov,  \tit{Noncommutative tori and universal
  sets of non-binary quantum gates}
  {\em Preprint} {\tt arXiv:quant-ph/0012009} (2000).
 \bibitem{VlaCla} A. Yu. Vlasov, \tit{Classical programmability is enough
  for quantum circuits universality in approximate sense}
  {\em Preprint} {\tt arXiv:quant-ph/0103119} (2001).
 \bibitem{VlaUQP} A. Yu. Vlasov, \tit{Universal quantum processors
  with arbitrary radix $n \ge 2$}
  {\em Preprint} {\tt arXiv:quant-ph/0103127} (2001).
\end{thebibliography}
\end{document}